\pdfoutput=1
\RequirePackage[l2tabu,orthodox]{nag}
\documentclass{PoS}
\usepackage[T1]{fontenc}
\usepackage[utf8]{inputenc}
\usepackage{microtype}
\usepackage{dsfont}
\usepackage{mathtools}
\usepackage{csquotes}
\usepackage{xfrac}
\usepackage{physics}
\usepackage{booktabs}
\usepackage[nolist]{acronym}
\usepackage{siunitx}
\usepackage{tikz}
\usepackage{pgf}

\usepackage[style=base]{caption}
\usepackage{subcaption}
\NewDocumentCommand\I{}{\mathrm{i}}
\NewDocumentCommand\e{ m }{\mathrm{e}^{#1}}
\NewDocumentCommand\hypgeo{ m m }{%
  \operatorname{%
    {\vphantom{\mathnormal{F}}}_{#1}%
    \kern-\scriptspace
    \mathnormal{F}_{#2}%
  }%
}
\NewDocumentCommand\thetaW{}{\theta_{\mathrm{W}}}
\NewDocumentCommand\sinW{}{\sin^2\thetaW}
\DeclareSIUnit\fm{\femto\metre}

\title{The leading hadronic contribution\\to the running of the Weinberg angle\\using covariant coordinate-space methods}

\ShortTitle{The leading hadronic contribution to $\sinW$ running using CCS methods}

\author{\speaker{Marco Cè}\\
       Helmholtz-Institut Mainz, Johannes Gutenberg-Universität Mainz, Germany\\
       E-mail: \email{marco.ce@uni-mainz.de}}

\author{Antoine Gérardin\\
       Helmholtz-Institut Mainz, Johannes Gutenberg-Universität Mainz, Germany\\
       E-mail: \email{gerardin@uni-mainz.de}}

\author{Konstantin Ottnad\\
       PRISMA Cluster of Excellence and Institut für Kernphysik, Johannes Gutenberg-Universität Mainz, Germany\\
       E-mail: \email{kottnad@uni-mainz.de}}

\author{Harvey B.\ Meyer\\
       PRISMA Cluster of Excellence and Institut für Kernphysik and Helmholtz-Institut Mainz, Johannes Gutenberg-Universität Mainz, Germany\\
       E-mail: \email{meyerh@uni-mainz.de}}

\abstract{%
We present a preliminary study of the leading hadronic contribution to the running of the Weinberg angle $\thetaW$.
The running is extracted from the correlation function of the electromagnetic current with the vector part of the weak neutral current using both the standard time-momentum representation method and the Lorentz-covariant coordinate-space method recently introduced by Meyer.
Both connected and disconnected contributions have been computed on $N_{\mathrm{f}}=2+1$ non-perturbatively $\order{a}$-improved Wilson fermions configurations from the \acl{CLS} initiative.
Similar covariant coordinate-space methods can be used to compute the leading hadronic contribution to the anomalous magnetic moment of the muon $(g-2)_\mu$ and to the running of the QED coupling $\alpha$.
\vspace*{0.5cm}
\begin{flushright}
MITP/18-098
\end{flushright}
}

\FullConference{The 36th Annual International Symposium on Lattice Field Theory - LATTICE2018\\
		22-28 July, 2018\\
		Michigan State University, East Lansing, Michigan, USA.}

\begin{acronym}
  \acro{HVP}{hadronic vacuum polarization}
  \acro{TMR}{time-momentum representation}
  \acro{CCS}{Lorentz-covariant coordinate space}
  \acro{FFT}{fast Fourier transform}
  \acro{CLS}{Coordinated Lattice Simulations}
  \acro{OBC}{open boundary condition}
\end{acronym}

\begin{document}

\section{Introduction}

The Weinberg angle or weak mixing angle $\thetaW$ is the parameter of the Standard Model of particle physics that parametrizes the mixing between electromagnetic and weak interactions
\begin{equation}
  \sinW = \frac{g'^2}{g^2+g'^2} , \qquad e = g \sin\thetaW = g' \cos\thetaW
\end{equation}
where $g$ and $g'$ are the $\mathrm{SU}(2)_L$ and $\mathrm{U}(1)_Y$ couplings, respectively.
As a consequence of the running with energy of $g$ and $g'$, the Weinberg angle is a function of the energy scale $Q^2$
\begin{equation}
  \sinW(Q^2) = \sinW \left[ 1 + \Delta\sinW(Q^2) \right] ,
\end{equation}
where $\sinW=\sinW(Q^2=0)=\num{0.23871(9)}$~\cite{Kumar:2013yoa} is the value in low-energy limit.
In particular, the leading hadronic contribution to the running is given by~\cite{Jegerlehner:1985gq,Jegerlehner:2011mw}
\begin{equation}
  \Delta_\text{had}\sin^2 \thetaW(Q^2) = -\frac{e^2}{\sinW} \left[ \Pi^{\gamma Z}(Q^2) - \Pi^{\gamma Z}(0) \right] ,
\end{equation}
proportional to the subtracted \ac{HVP}
\begin{equation}
  (Q_\mu Q_\nu-\delta_{\mu\nu}Q^2)\Pi^{\gamma Z}(Q^2) = \Pi_{\mu\nu}^{\gamma Z}(Q^2) = \int\dd[4]{x} \e{\I Q\cdot x} \ev{j_\mu^Z(x) j_\nu^\gamma(0)}
\end{equation}
of the electromagnetic current $j_\mu^\gamma$ and the vector part of the neutral weak current $j_\mu^Z$
\begin{subequations}
  \begin{gather}
    j_\mu^\gamma = \frac{2}{3}\bar{u}\gamma_\mu u - \frac{1}{3}\bar{d}\gamma_\mu d - \frac{1}{3}\bar{s}\gamma_\mu s + \frac{2}{3}\bar{c}\gamma_\mu c , \\
    j_\mu^Z      = j_\mu^3 - \sinW j_\mu^\gamma , \qquad
    j_\mu^3      = \frac{1}{4}\bar{u}\gamma_\mu u - \frac{1}{4}\bar{d}\gamma_\mu d - \frac{1}{4}\bar{s}\gamma_\mu s + \frac{1}{4}\bar{c}\gamma_\mu c .
  \end{gather}
\end{subequations}
The vacuum polarization $\Pi^{\gamma Z}$ is directly accessible to lattice computations~\cite{Burger:2015lqa,Guelpers:2015nfb}.
The computation is similar to that of the leading \ac{HVP} contribution to the anomalous magnetic moment of the muon and, as in that case, different methods are available, such as the four-momentum approach or the \ac{TMR} method~\cite{Bernecker:2011gh}.
Using the latter, $\Pi^{\gamma Z}$ is given by
\begin{equation}
  \Pi^{\gamma Z}(Q^2) - \Pi^{\gamma Z}(0) = \int_0^\infty \dd{x_0}  G^{\gamma Z}(x_0) K(x_0,Q^2) , \qquad K(x_0,Q^2) = x_0^2-\frac{4}{Q^2}\sin[2](\frac{Qx_0}{2}) ,
\end{equation} 
where $G^{\gamma Z}(x_0)$ is the zero-momentum projection of the correlator
\begin{equation}
  G^{\gamma Z}(x_0) = -\frac{1}{3}\sum_{k=1,2,3} \int\dd[3]{x} G^{\gamma Z}_{kk}(x) , \qquad G^{\gamma Z}_{\mu\nu}(x) = \ev{j_\mu^Z(x) j_\nu^\gamma(0)} .
\end{equation}
Writing explicitly Wick's contractions results in both connected and disconnected contributions
\begin{multline}
  G^{\gamma Z}_{\mu\nu}(x) = \left(\frac{1}{ 4}-\frac{5}{9}\sinW\right) C^{\ell,\ell}_{\mu\nu}(x)
                           + \left(\frac{1}{12}-\frac{1}{9}\sinW\right) C^{s,s}_{\mu\nu}(x) \\
                           + \left(\frac{1}{ 6}-\frac{4}{9}\sinW\right) C^{c,c}_{\mu\nu}(x)
                           -                    \frac{1}{9}\sinW        D^{\ell+As,\ell-s}_{\mu\nu}(x) ,
\end{multline}
where $A=3/(4\sinW)-1$, the disconnected charm contribution has been neglected, and
\begin{subequations}
  \begin{gather}
    C^{f_1,f_2}_{\mu\nu}(x) = -\ev{\Tr{D_{f_1}^{-1}(x,0)\gamma_\mu     D_{f_2}^{-1}(0,x)\gamma_\nu}} ,\\
    D^{f_1,f_2}_{\mu\nu}(x) =  \ev{\Tr{D_{f_1}^{-1}(x,x)\gamma_\mu}\Tr{D_{f_2}^{-1}(0,0)\gamma_\nu}} .
  \end{gather}
\end{subequations}

\section{The \acs{CCS} method}

As an alternative to the \ac{TMR} method, in this work we also implement the recently proposed \ac{CCS} method~\cite{Meyer:2017hjv}.
We rewrite the subtracted vacuum polarization as a covariant integral in four-dimensional coordinate space
\begin{gather}
  \Pi^{\gamma Z}(Q^2) - \Pi^{\gamma Z}(0) = \int\dd[4]{x} G^{\gamma Z}_{\mu\nu}(x) H_{\mu\nu}(x) = \smashoperator[l]{\sum_{i=1,2}} \int\dd[4]{x} g_i(x) \mathcal{H}_i(\abs{x}) , \label{eq:hvp_ccs}\\
  g_1(x) = -\delta_{\mu\nu}G^{\gamma Z}_{\mu\nu}(x) , \qquad g_2(x) = \frac{x_\mu x_\nu}{x^2}G^{\gamma Z}_{\mu\nu}(x) ,
\end{gather}
where we used the Lorentz structure of the \ac{CCS} kernel
\begin{equation}
\label{eq:kernel_ccs}
  H_{\mu\nu}(x) = -\delta_{\mu\nu} \mathcal{H}_1(\abs{x}) + \frac{x_\mu x_\nu}{x^2} \mathcal{H}_2(\abs{x}) , \qquad \mathcal{H}_i(\abs{x}) = x^2 \bar{\mathcal{H}}_i(\abs{Q}\abs{x}) ,
\end{equation}
that can be expressed in term of generalized hypergeometric functions $\hypgeo{2}{3}$
\begin{subequations}
\begin{gather}
  \begin{multlined}
    \bar{\mathcal{H}}_1(z) = \frac{z^2}{4608} \left[ 24\hypgeo{2}{3}\left(1,1;2,3,3;-z^2/4\right) -20\hypgeo{2}{3}\left(1,1;2,3,4;-z^2/4\right) \right.\\
                                              \left. +3\hypgeo{2}{3}\left(1,1;2,3,5;-z^2/4\right) \right] ,
  \end{multlined}\\
  \begin{multlined}
    \bar{\mathcal{H}}_2(z) = \frac{z^2}{1152} \left[  6\hypgeo{2}{3}\left(1,1;2,3,3;-z^2/4\right) - 8\hypgeo{2}{3}\left(1,1;2,3,4;-z^2/4\right) \right.\\
                                              \left. +4\hypgeo{2}{3}\left(1,1;2,4,4;-z^2/4\right) -  \hypgeo{2}{3}\left(1,1;2,4,5;-z^2/4\right) \right] .
  \end{multlined}
\end{gather}
\end{subequations}

\subsection{Non-transverse kernel}

The kernel defined in Eq.~\eqref{eq:kernel_ccs} is transverse, i.e.\ it satisfies the condition $\partial_\mu H_{\mu\nu}(x)=0$, or equivalently $\abs{x}\mathcal{H}'_1(\abs{x})=\abs{x}\mathcal{H}'_2(\abs{x})+3\mathcal{H}_2(\abs{x})$.
We can modify it by adding a non-transverse component
\begin{equation}
  \partial_\mu [x_\nu \mathcal{F}(\abs{x})] = \delta_{\mu\nu}\mathcal{F}(\abs{x}) + \frac{x_\mu x_\nu}{x^2} \abs{x}\mathcal{F}'(\abs{x}) .
\end{equation}
Using the fact that $\partial_\mu G_{\mu\nu}(x)=0$, this modification entails only a surface term $\int\dd[4]{x}\partial_\mu[G_{\mu\nu}x_\nu \mathcal{F}]$, that vanishes in infinite volume.
We parametrize the non-transverse component with $\mathcal{F}(\abs{x})=(\alpha+\gamma)\mathcal{H}_1(\abs{x})-\gamma\mathcal{H}_2(\abs{x})$, so that the following choices of the $\alpha$ and $\gamma$ parameters
\begin{equation}
  H_{\mu\nu}(x) = \begin{cases}
    -\delta_{\mu\nu} \left[ 4\mathcal{H}_1(\abs{x}) - \mathcal{H}_2(\abs{x}) \right]/3 ,             & \alpha=0,\, \gamma=-1/3 , \\
    -\delta_{\mu\nu} \mathcal{H}_2(\abs{x}) + 4\frac{x_\mu x_\nu}{x^2} \mathcal{H}_2(\abs{x}) ,      & \alpha=0,\, \gamma=1 , \\
    \frac{x_\mu x_\nu}{x^2} \left[ \mathcal{H}_2(\abs{x}) + \abs{x}\mathcal{H}'_1(\abs{x}) \right] , & \alpha=1,\, \gamma=0 ,
  \end{cases}
\end{equation}
correspond respectively to kernels that have a \enquote{monopole} $\delta_{\mu\nu}$ structure, a traceless \enquote{quadrupole} one, and a $(x_\mu x_\nu)/x^2$ structure.
Depending on the value of $\alpha$ and $\gamma$, the kernel samples differently the correlator at short or long distances.
In calculating $\Pi'(0)$, one easily shows that, using a kernel containing no $\delta_{\mu\nu}$ tensor structure, the integrand falls off at long distances with one power of $\abs{x}$ faster than with the transverse kernel, if one assumes the correlator to be dominated by a single vector meson.
In our case, both the $\alpha=1$ and $\gamma=1$ options result in shorter-range kernels, which in turn leads to smaller statistical errors from the integration tail, but might be affected by larger short-distance discretization artefacts.

\subsection{Lattice discretization and $\order{a}$ improvement}

The lattice discretization of Eq.~\eqref{eq:hvp_ccs} is straightforward and amounts to just employing the lattice-determined $G^{\gamma Z}_{\mu\nu}(x)$.
Here we show only the $\order{a}$-improved local vector current
\begin{equation}
  (V_{\mathrm{I}})_\mu(x) = V_\mu(x) + a c_V \partial_\alpha T_{\mu\alpha}(x) , \qquad V_\mu(x)=\bar{\psi}(x)\gamma_\mu\psi(x), \quad T_{\mu\alpha}(x) = -\frac{1}{2}\bar{\psi}(x)\comm{\gamma_\mu}{\gamma_\alpha}\psi(x) ,
\end{equation}
with non-perturbatively determined improvement coefficient $c_V$ and renormalization~\cite{Bhattacharya:2005rb,Gerardin:2018kpy}, but the conserved definition has also been studied.
Improving the integral in Eq.~\eqref{eq:hvp_ccs} results in
\begin{multline}
  \int\dd[4]{x} \left\{ \ev{V_\mu(x)V_\nu(0)} H_{\mu\nu}(x) - ac_V \left[ \ev{T_{\mu\alpha}(x) V_\mu(0)} - \ev{V_\mu(x) T_{\mu\alpha}(0)} \right] \partial_\mu H_{\mu\nu}(x) \right\} \\
  = \smashoperator[l]{\sum_{i=1,2,3}} \int\dd[4]{x} g_i(x) \mathcal{H}_i(\abs{x}) ,
\end{multline}
where we integrated by part and used translation invariance, and the $\order{a}$-improvement contribution is encoded in the $g_3$ correlator and the $\mathcal{H}_3$ kernel
\begin{equation}
  g_3(x) = \frac{a x_\alpha}{x^2} c_V \left[ \ev{T_{\mu\alpha}(x) V_\mu(0)} - \ev{V_\mu(x) T_{\mu\alpha}(0)} \right] , \qquad \mathcal{H}_3(\abs{x}) = \abs{x} \mathcal{H}'_1(\abs{x}) + \mathcal{H}_2(\abs{x}) .
\end{equation}

\section{Numerical tests}

\begin{table}[b]
  \centering\small
  \caption{Parameters and number of configurations of the \acs{CLS} ensemble used.}\label{tab:ensembles}
  \begin{tabular}{rS[table-format=2]S[table-format=1.1]S[table-format=3]S[table-format=3]S[table-format=1.1]S[table-format=4]S[table-format=4]S[table-format=3]S[table-format=3]}
    \toprule
    & {$L/a$} & {$L$ [\si{\fm}]} & {$m_\pi$ [\si{\MeV}]} & {$m_K$ [\si{\MeV}]} & {$m_\pi L$} & \multicolumn{4}{c}{\#cnfg ($\ell$, $s$, $c$, disc.)} \\
    \midrule
    N203 & 48 & 3.1 & 340 & 440 & 5.4 & 1504 &  752 &  94 & 752 \\
    N200 & 48 & 3.1 & 280 & 460 & 4.4 & 1712 &  856 & 107 & 856 \\
    D200 & 64 & 4.1 & 200 & 480 & 4.2 & 1080 & 1080 & 135 & 270 \\
    \bottomrule
  \end{tabular}
\end{table}

\begin{figure}[t]
  \resizebox{\columnwidth}{!}{\input{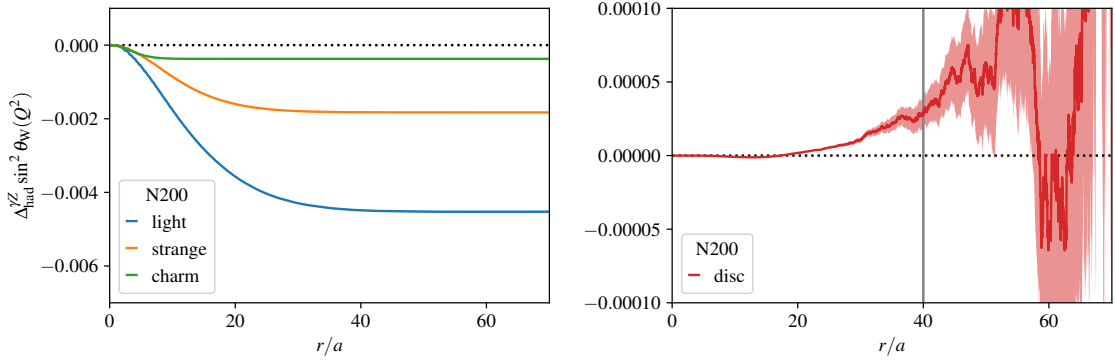}}
  \caption{Plots of the connected (left) and disconnected (right) contributions to $\Delta_\text{had}\sinW(Q^2)$ at a scale $Q^2=\SI{4}{\GeV\squared}$ with respect to the integration cut $r$, using the CCS method on the N200 ensemble. In the disconnected case, the chosen value of $r$ is indicated by a vertical line.}\label{fig:contributions}
\end{figure}

We tested the method on three $N_{\mathrm{f}}=2+1$ ensembles from the \ac{CLS} initiative~\cite{Bruno:2014jqa} listed in Table~\ref{tab:ensembles}.
On all three ensembles, $\beta=3.55$ and $a\simeq\SI{0.065}{\fm}$~\cite{Bruno:2016plf}.
The light, strange and valence-charm connected contributions have been computed performing inversions on $5$ ($10$-$15$ for the light) point sources randomly placed in space per configuration.
The light and strange disconnected contributions have been computed estimating the trace of the quark propagator $L_\mu(y)=\Tr{D^{-1}(y,y)\gamma_\mu}$ using \num{2} random sources of \num{512} hierarchical probing vectors~\cite{Stathopoulos:2013aci}.
Disconnected all-to-all two-point functions are computed efficiently implementing the correlation of the two stochastically-estimated traces using the \ac{FFT}
\begin{equation}
  D_{\mu\nu}(x) = \int\dd[4]{y} \ev{L_\mu(y) L_\nu(y+x)} = \int \dd[4]{x} \e{-\I p\cdot x} \ev{\hat{L}_\mu(-p) \hat{L}_\nu(p)} , \quad \hat{L}_\mu(p) = \int \dd[4]{y} \e{\I p\cdot y} L_\mu(y) ,
\end{equation}
with suitable modifications to handle the \aclp{OBC} in the time direction correctly.

\subsection{Integration strategy}

In order to implement the four-dimensional integral in Eq.~\eqref{eq:hvp_ccs} on the lattice, we define contributions summed over lattice points in a four-dimensional sphere of radius $r$
\begin{equation}
  \Delta_\text{had}\sin^2 \thetaW(Q^2,r) = -\frac{e^2}{\sinW} \int_{S_{r,L}} \dd[4]{x} s(r,L) g_i(x) \mathcal{H}_i(\abs{x}) , \quad S_{r,L} = \{ x : \abs{x}<r ,\, x_i<L/2-\delta \} .
\end{equation}
Having to deal with lattices of finite size $L$ in space directions, we modify the sum including only points that satisfy $x_i<L/2-\delta$ for $i=1,2,3$, with $\delta=4a$, and correcting for the missing points with an exactly calculable geometric factor $s(r,L)$.

The different contributions are plotted in Figure~\ref{fig:contributions} against the value of $r$.
The left plot shows that the light, strange and valence-charm contribution are precisely determined, with statistical errors that are smaller than the line thickness.
At large $\abs{x}$, the exponential suppression of both the signal and the statistical error of $g_i(x)$ dominates over the polynomial growth of the kernel, and the connected contributions can be extracted for $r\to\infty$.

\subsection{Long-distance systematics of the disconnected contribution}

In contrast with the connected-contribution case, the statistical error of the disconnected integrand grows at large $\abs{x}$.
As a consequence, the integral in the right plot in Figure~\ref{fig:contributions} looses its signal at distances $\gtrsim 50a$.
To estimate the disconnected contribution, we cut the integration of the noisy correlator tail.
In order to constrain the induced systematic effect, we observe that the long-distance behaviour of $G^{\gamma Z}(x)$ is dominated by the isospin-triplet component
\begin{equation}
  G^{\gamma Z}(x) \stackrel{x\to\infty}{\sim} \left(\frac{1}{2}-\sinW\right) G^{I=1}(x) , \qquad G^{I=1}(x) = \frac{1}{2} C^{\ell,\ell}(x) .
\end{equation}
This component is easily estimated since it does not include disconnected contributions. In turns, the ratio between the disconnected contribution $G^{\gamma Z}_{\text{disc}}(x)$ and $G^{I=1}(x)$ tends asymptotically to a constant
\begin{multline}
  \frac{G^{\gamma Z}_{\text{disc}}(x)}{G^{I=1}(x)} = \frac{G^{\gamma Z}(x)-\left(\frac{1}{2}-\sinW\right) G^{I=1}(x)}{G^{I=1}(x)} + \frac{1}{9}\sinW \\
  - \left(\frac{1}{6}-\frac{2}{9}\sinW\right) \frac{C^{s,s}(x)}{C^{\ell,\ell}(x)} - \left(\frac{1}{3}-\frac{8}{9}\sinW\right) \frac{C^{c,c}(x)}{C^{\ell,\ell}(x)}\stackrel{x\to\infty}{\sim} \frac{1}{9}\sinW.
\end{multline}
Thus, we compare the statistical error of $\Delta_\text{had}^\text{disc}\sin^2 \thetaW(Q^2,r)$ with the value of $\Delta_\text{had}^{I=1}\sin^2 \thetaW(Q^2,r\to\infty)-\Delta_\text{had}^{I=1}\sin^2 \thetaW(Q^2,r)$.
The latter represents an upper bound on the neglected contribution, and $r$ is chosen such that this upper bound is less than half of the statistical error of $\Delta_\text{had}^\text{disc}\sin^2 \thetaW(Q^2,r)$.

\section{Conclusions}

\begin{figure}[t]
  \resizebox{\columnwidth}{!}{\input{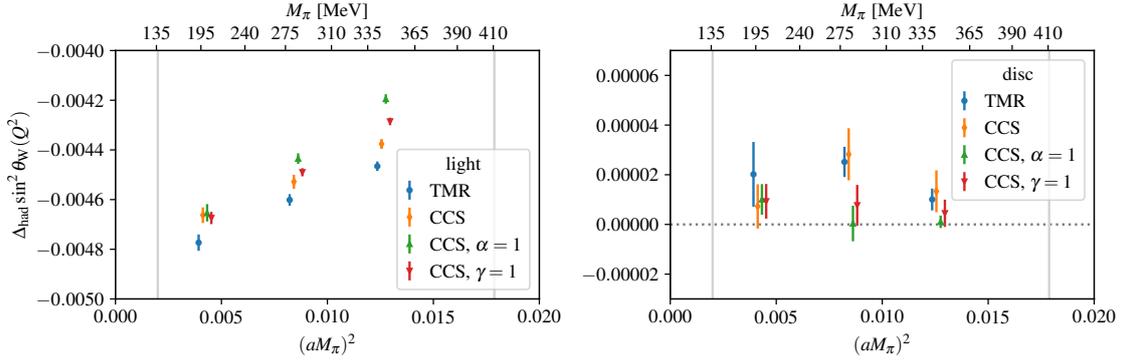}}
  \caption{Preliminary light connected (left) and disconnected (right) contributions on all ensembles and with different methods. The left vertical line indicates the physical pion and kaon masses point, the right one indicates the $\mathrm{SU}(3)$ flavour symmetric point.}\label{fig:results}
\end{figure}

\begin{table}[b]
  \centering\small
  \caption{Preliminary value of $\Delta_\text{had}\sinW(Q^2)$ at $Q^2=\SI{4}{\GeV\squared}$ and its connected (light, strange and charm) and disconnected contributions, on three different ensembles using the \ac{TMR} and the \ac{CCS} methods.}\label{tab:results}
  \begin{tabular}{rS[table-format=+4(2)]S[table-format=+4(2)]S[table-format=+4(2)]S[table-format=+4(2)]S[table-format=+4(2)]S[table-format=+4(2)]}
    \toprule
    $\times\:\num{e-6}$ & \multicolumn{2}{c}{N203 (TMR, CCS)} & \multicolumn{2}{c}{N200 (TMR, CCS)} & \multicolumn{2}{c}{D200 (TMR, CCS)} \\
    \midrule
    light   & -4466(19) & -4376(20) & -4601(23) & -4528(28) & -4773(33) & -4663(32) \\
    strange & -1961(7)  & -1901(6)  & -1888(6)  & -1828(5)  & -1757(4)  & -1689(3)  \\
    charm   &  -443(2)  &  -364(1)  &  -451(2)  &  -373(1)  &  -455(1)  &  -377(1)  \\
    disc    &    10(4)  &    13(8)  &    25(6)  &    28(10) &    20(13) &     8(10) \\
    \midrule
    total   & -6859(22) & -6628(24) & -6916(26) & -6700(30) & -6965(35) & -6721(34) \\
    \bottomrule
  \end{tabular}
\end{table}

Preliminary results in Table~\ref{tab:results} and Figure~\ref{fig:results} indicate that we are able to estimate the disconnected contribution with a statistical error that is only \num{0.1}-\SI{0.2}{\percent} of $\Delta_\text{had}\sinW(Q^2)$ at $Q^2=\SI{4}{\GeV\squared}$, and the systematic errors from the truncation of the spacetime summation under control.
The statistical error is dominated by the light connected contribution, that has been computed with a low statistics and can easily be improved.
A direct comparison between the \ac{TMR} and \ac{CCS} methods has been performed:
In the connected-contribution case, the two methods result in the same statistical precision, and the difference in the central values can be attributed to different discretization effects.
In the disconnected-contribution case, the statistical precision of the \ac{CCS} method shows a different volume dependence than the \ac{TMR} method one.
In particular, disconnected contributions estimated with the \ac{CCS} method are comparatively slightly less precise on the two $(\SI{3.1}{\fm})^3$ boxes, but they are more precise on the larger-volume ensemble, in particular when a non-transverse kernel is employed.
This suggests to further investigate \ac{CCS} methods on lattices with a large physical volume, applying them also to the computation of the leading \ac{HVP} contribution to $(g-2)_\mu$ and to the running of the QED coupling $\alpha$.
Varying the energy scale up to $Q^2=\SI{10}{\GeV\squared}$ does not affect these conclusions.

The results presented in Table~\ref{tab:results} are preliminary and a full assessment of systematic errors is lacking.
The main systematics to be assessed are finite-volume effects and scale-setting errors.
Moreover, ensembles with different lattice spacings are needed to confirm that different methods result in compatible values in the continuum limit.
Nevertheless, these results shows that it is possible to achieve a sub-percent determination of $\Delta_\text{had}\sinW(Q^2)$, including the disconnected contribution with full control of the integration tail systematics.
In general, lattice methods compare very favourably to the phenomenological determination~\cite{Jegerlehner:2011mw}, which is affected by the systematics from flavour separation.

{\footnotesize
\textbf{Acknowledgements:}
Calculations for this project have been performed on the HPC clusters \enquote{clover} and \enquote{himster2} at Helmoltz-Institut Mainz and \enquote{Mogon II} at JGU Mainz, and on the BG/Q system \enquote{JUQUEEN} at JSC, Jülich.
The authors gratefully acknowledge the support of the John von Neumann Institute for Computing and the Gauss Centre for Supercomputing for project HMZ21.
We are grateful to our colleagues in the \ac{CLS} initiative for sharing ensembles.
}

\bibliographystyle{JHEP}
\bibliography{./biblio.bib}

\end{document}